\documentclass[twoside,british,english]{elsarticle}
\usepackage[T1]{fontenc}
\usepackage[latin9]{inputenc}
\usepackage{textcomp}
\usepackage{amsbsy}
\usepackage{amstext}
\usepackage{graphicx}
\usepackage{esint}

\makeatletter
\journal{arXiv.org Cornell University}

\makeatother

\usepackage{babel}
\begin{document}

\begin{frontmatter}{}

\title{On Stress Driven Diffusion in Bone - An Experimental Study}

\author{Gustav Lindberg$^{a}$and Per Ståhle$^{b}$}

\address{$^{a}$\foreignlanguage{british}{Solid Mechanics, Dept. of Eng. Mech.,
KTH Royal Institute of Tech. Stockholm, Sweden, }}

\address{$^{b}$Solid Mechanics, Lund Institute of Technology, Lund University,
Lund, Sweden\foreignlanguage{english}{,}}
\begin{abstract}
The transport of nutrients or signal constituents that stimulate growth
of bone tissue is supposed to be affected by a static mechanical load.
It follows from basic thermodynamical principles that constituents
causing volumetric change are dragged along the gradients of hydrostatic
stress. The present preliminary study examines the behaviour of iodine
present in the medullary cavity of a bovine long bone exposed to mechanical
load. A section of the bone is x-ray scanned with the static load
present, with and without the iodine. The resulting distribution in
a selected 2D plane is numerically evaluated using a discrete Radon's
inverse transform. The result suggests that iodine is a useful constituent
with a good attenuation effect on the x-ray beam and clearly reveals
the temporal distribution of its transport through the bone. It further
result shows some indication that stress does affect the iodine distribution.
\foreignlanguage{english}{.}
\end{abstract}

\end{frontmatter}{}

\section{Introduction}

Bone alters its shape and structure when subjected to mechanical loading
{[}1-3{]}. The hollow cross-sections of long bones offer a density
effective mean for supporting a large variety of loading modes. It
has been shown that long bones grow in places where the stress is
large to form a cross-section that increases its load carrying capacity
and by decreasing the largest cross-sectional stress {[}4-6{]}.

Low physical activity, e.g., bed rest or being under low-gravity conditions,
leads to bone tissue resorption, whereas increased activity stimulates
bone growth. Moreover, it has been observed that static loading does
not cause bone modelling {[}7{]}. From a mechanical perspective the
most effective way to strengthen a long bone is to add bone at the
outer surface, i.e., the periosteal surface. This increases the elastic
section modulus and therefore makes it less prone to fracture. 

The present investigation, is motivated by the proposal that the modelling
process is governed by changes in the chemical environment at the
periosteal surface. The hypothesis focuses on molecular transport
in solids, driven by the gradients of the mechanical stress, as derived
by Einstein and observed by Smoluchowski {[}8, 9{]}. In the present
case the transport from the medullary cavity to the periosteal surface
is considered. In the present pilot study iodine is selected as a
replacement substance for its large cross-sectional optical area leading
to a significant scattering and attenuation of the x-ray beam.

\section{Samples and experiment}

The x-ray scanning was performed at The Henry Moseley X-ray Imaging
Facility, Photon Science Institute at the University of Manchester.
The scanned sample of an around 36\textpm 2.5mm high bovine long bone
with an ellipsoidal cross-section with 39mm and 53mm across. Fig.
1 shows the starting points for two scan series made first without,
and then with, an iodine solution introduced into the medullary cavity.
The crack that is visible in the figure is opened by a plastic wedge.
A silicon adhesive was used for sealing the opened crack to prevent
leakage. The same adhesive was used to attach the bone to a polycarbonate
plate which was used to support the bone.

Altogether 6284 scans were made at equiangular steps from -180 to
180 degrees for each of the two tests. Each image have a resolution
of 4000 times 4000 pixels using 16 bits representing the grayscale.
The duration of a full scan was a couple of minutes. The duration
between the first test without and the second with the iodine solution
was close to 30 minutes. 

\begin{figure}
\begin{raggedright}
a) \hspace{0.4\textwidth}b)\\
\foreignlanguage{british}{\includegraphics[width=0.48\textwidth]{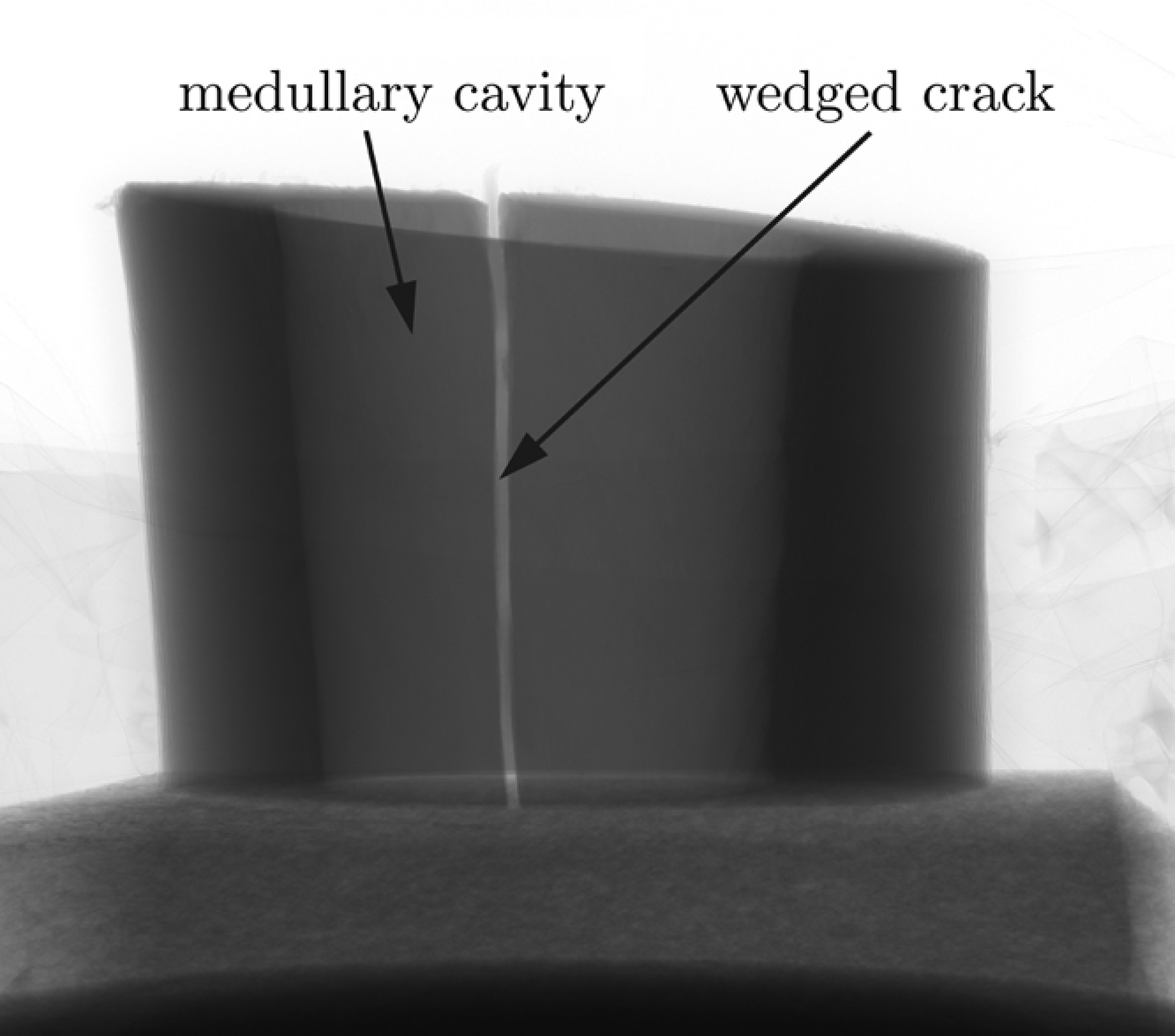}
\includegraphics[width=0.48\textwidth]{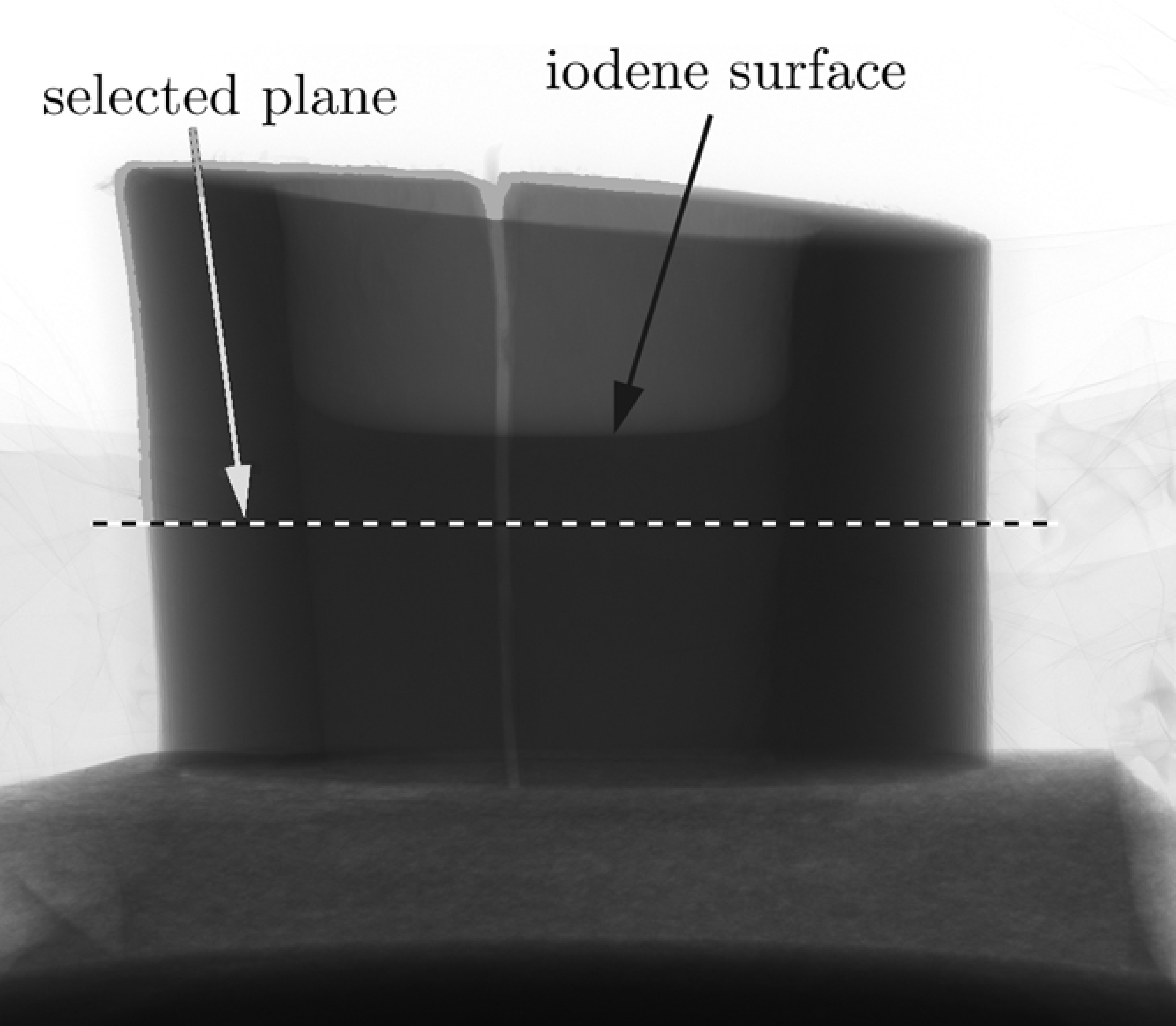}}
\par\end{raggedright}
\centering{}\label{Fig:boneSamples}\caption{Bone sample a) without iodine, and b) with iodine inserted into the
medullary cavity. The iodine volume is the darkened part below the
marked iodine surface.}
\end{figure}

\section{Reconstruction of a plane cut through a body}

Consider a finite body that is exposed to impinging collimated light,
cf. Fig. 2. The interaction between the body and the light is assumed
to be restricted to attenuation of the light only, while it stays
collimated. Two cartesian coordinate systems $x$-$y$ and $\xi$-$\eta$
with common origin are attached to a plane cross-section of the body.
The $\xi$ axis is rotated the angle $\theta$ from the $x$-axis
as is shown in Fig. 2. A polar coordinate system is defined as $\rho=\sqrt{x^{2}+y^{2}}$
and $\theta=\arctan(y/x$).

\begin{figure}
\begin{centering}
\includegraphics[width=0.45\textwidth]{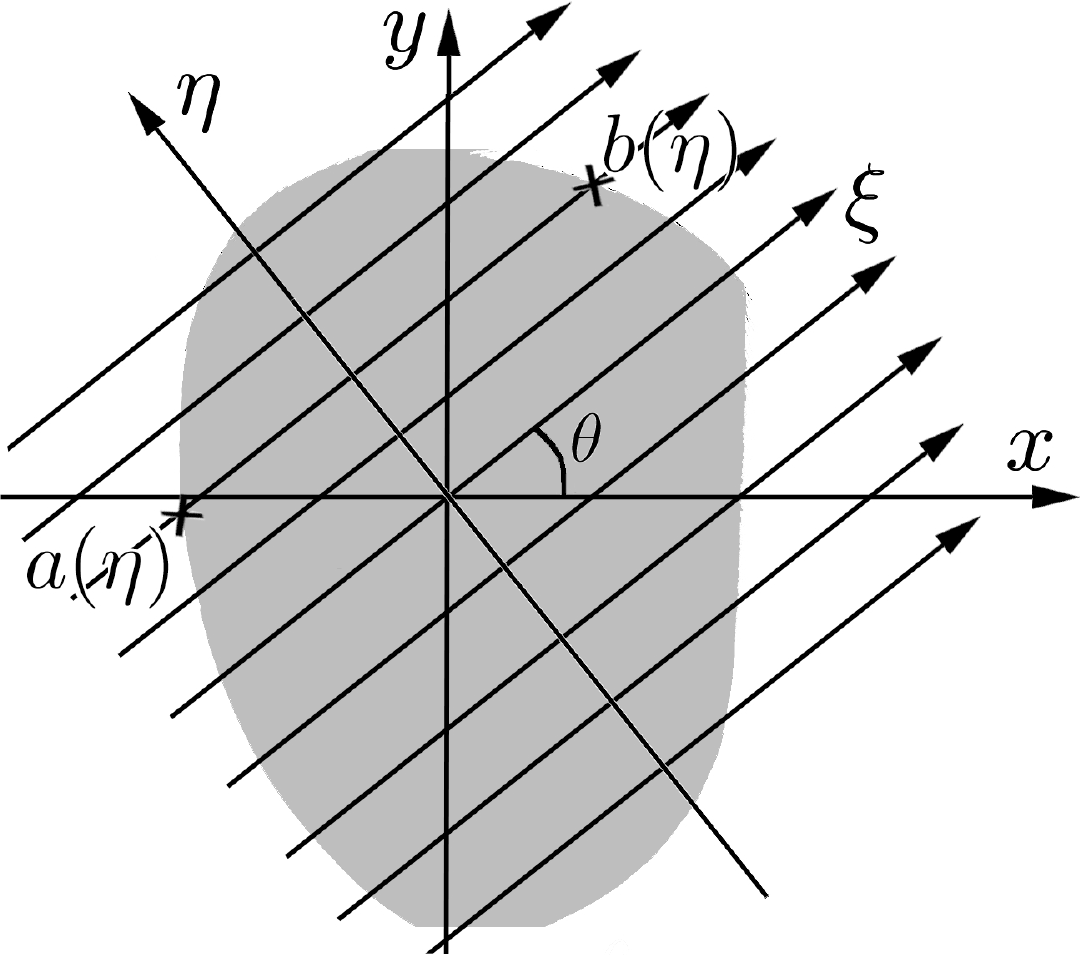}
\par\end{centering}
\centering{}\label{Fig:coordinates}\caption{Light impinging on a semitransparent body. The $a(\eta)$ and $b(\eta)$
mark the entrance and exit of the x-ray.}
\end{figure}

Consider a single beam of light that transverses the body in the positive
$\xi$-direction, i.e., at a constant $\eta$. Along its path the
light intensity $I=I(\xi,\eta)$ becomes attenuated at a rate $\text{d}I/\text{d}\xi$.
The optical density $\mu$ of the material causes the attenuation
and is supposed to vary in the plane. The attenuation rate is also
assumed to be proportional to the intensity $I$ and therefore, 

\begin{equation}
\text{d}I(\xi,\eta)/\text{d}\xi=-\mu(\xi,\eta)I(\xi,\eta).
\end{equation}
Integration readily gives the optical density as,

\begin{equation}
\mu(\xi,\eta)=-\frac{\text{d}}{\text{d}\xi}\ln I(\xi,\eta).\label{eq:mu}
\end{equation}
The light crosses the body boundary twice, i.e., at entrance $\xi_{a}=a(\eta)$
and when it leaves the body at $b(\eta)$, see Fig. 2. Integration
is performed along the $\xi$-axis from the entry, where the impinging
intensity, $I_{o}$, is assumed to be independent of $\eta$. After
the exit of the body at $\xi_{b}=b(\eta)$ the intensity remains independent
of $\xi$, i.e., from (\ref{eq:mu}) we get

\begin{equation}
I(\xi,\eta)/I_{o}=\exp[\intop_{a}^{\xi}\mu(\xi',\eta)\text{d}\xi']=\left\{ \begin{array}{c}
1\quad\text{for}\quad\xi\le a(\eta)\\
\exp[\intop_{a}^{\xi}\mu(\xi',\eta)\text{d}\xi']\quad\text{for}\quad a(\eta)<\xi<b(\eta)\\
\exp[\intop_{a}^{b}\mu(\xi',\eta)\text{d}\xi']\quad\text{for}\quad\xi\le b(\eta)
\end{array}\right..
\end{equation}
The light intensity $I(\eta)$ remaining after passing the body at
$\xi=b$ is given by

\begin{equation}
\ln[I(\eta)/I_{o}]=f(b,\eta)-f(a,\eta),\quad\text{where}\quad f(\eta)=\intop_{a}^{b}\mu(\xi,\eta)\text{d}\xi,
\end{equation}

Consider now, multiple exposures as described above but now from different
angles. For the full exposure of the body let the function $g=g(\rho,\theta)$
be defined as follows

\begin{equation}
g(\rho,\theta)=\intop_{A}\mu(x,y)\delta(x\cos\theta+y\sin\theta-\rho)\text{d}x\text{d}y,
\end{equation}
where $\delta$ is Dirac's delta function. Assume further that $\mu(x,y)$
is unknown and that $g(\rho,\theta)$ is a known function. The Radon
inverse transformation is closely related to the inverse Fourier transform
and is given by {[}10, 11{]} as 

\begin{equation}
\mu(x,y)=\frac{1}{2\pi^{2}}\intop_{-\infty}^{\infty}\intop_{0}^{\pi}\frac{\partial g(\rho,\theta)}{\partial\rho}\frac{\text{d}\rho\text{d}\theta}{x\cos\theta+y\sin\theta-\rho}.\label{eq:InvRadon}
\end{equation}
With the known projection $g(\rho,\theta)$ the solution is readily
given through numerical integration of (\ref{eq:InvRadon}). With
the irregular geometry of the bone cross-section one cannot hope for
an analytical solution. However, the close connection to the Fourier
transform with a developed formalism for obtaining discretising solutions
for both the transform and its corresponding inverse transform, can
be utilised as is described in the following section. 

\section{Numerical analysis}

The numerical approximation of the optical density $\mu(x,y)$ relies
on the accuracy and resolution of the measured intensity function
$g(\rho,\theta)$. By discretising on the pixel level, a resulting
accurate density $\mu(x,y)$ is anticipated. The discretised integral
approximating (\ref{eq:InvRadon}) may be written

\begin{equation}
\mu(x,y)=\{\Re^{-1}g\}(x,y).\label{eq:RadonOper}
\end{equation}
by organising the discrete pixel intensities representing the functions
$\mu(x,y)$ and $g(\rho,\theta)$ in vectors $\boldsymbol{\mu}$ and
$\boldsymbol{g}$. We may write (\ref{eq:RadonOper}) as a system
of equations

\begin{equation}
\text{\ensuremath{\boldsymbol{\mu}}}=\text{\ensuremath{\boldsymbol{A}}}\text{\ensuremath{\boldsymbol{g}}}\label{eq:RadonNum}
\end{equation}
where $\boldsymbol{\mu}$ and $\boldsymbol{g}$ have the dimension
$N$ and hence \textbf{\textit{A}} is an $N\times N$ matrix. The
basic images $\left\{ h_{j}(x,y)\right\} _{j=1}^{J}$ are defined
by

\begin{equation}
h_{j}(x,y)=\left\{ \begin{array}{l}
1\text{ if }(x,y)\in\text{ pixel no. }j\\
0\text{ otherwise}
\end{array}\right.,
\end{equation}
and the image is written

\begin{equation}
g(x,y)=\sum_{j=1}^{J}c_{j}h_{j}(x,y),
\end{equation}
where $J$ is the number of pixels. The coefficients $c_{j}$ can
be determined provided that the number of projections is equal or
larger than the number of pixels, $J$. A more detailed description
of the numerics that come with the discrete inverse Radon Transform
is given in {[}10{]}.

\section{Results and discussion}

In the present study a total of 20 planes below and above a selected
cross-section as shown in Fig. \ref{Fig:boneSamples}b) are averaged
and used for the reconstruction of the optical density $\mu(x,y)$
represented as a 2D image.

\begin{figure}
\begin{centering}
\includegraphics[width=0.45\textwidth]{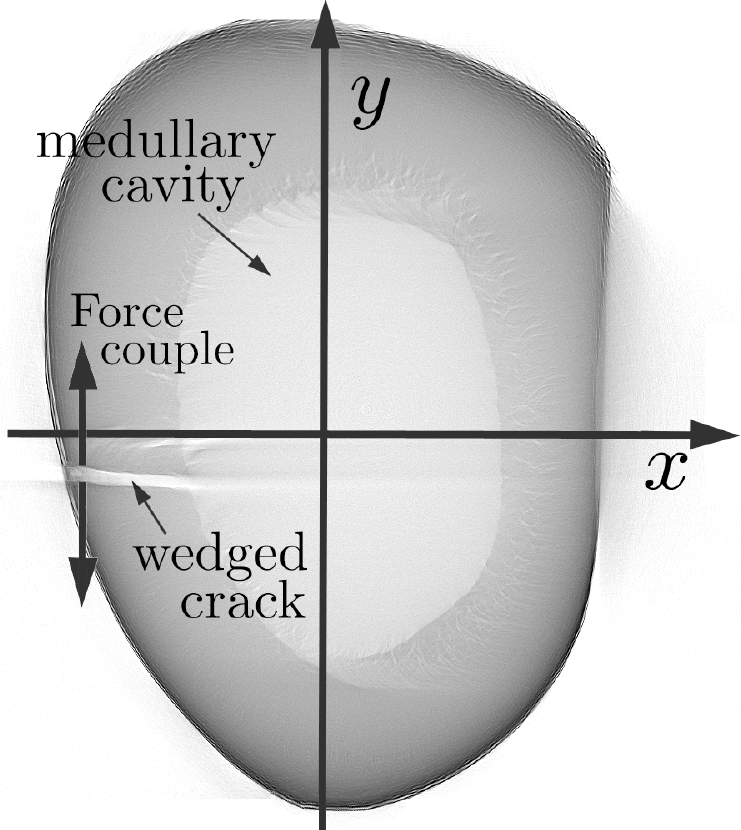}
\par\end{centering}
\centering{}\label{Fig:RadonRes-1}\caption{The obtained density field of the selected cross-section. The studied
densities are at $x=0$ and at $y=0$. The crack is the white cut
vaguely visual near the negative $x$-axis. The contrasting density
that vanishes in the empty opened crack creates an optical ghost mark
that is seen stretching into the medullary cavity.}
\end{figure}

Cartesian coordinates $x$ and $y$ are chosen along the shorter,
$x$, respectively the longer, $y$, axes of the ellipsoidal cross-section
(cf. Fig. 3). The initial incident x-rays are parallel with the $x$-axis.
The approach is to rotate the specimen 360 degrees around an axis
perpendicular to the $x$-$y$ plane. During a couple of minutes 6284
scans are made. 

Each scan gives a line of intensities such as the marked line in the
image taken at the first scan, i.e. Fig. \ref{Fig:boneSamples}. All
6284 give a 1D view of the selected plane from different angles 0
to 360 degrees. Together these define the function $g(x,y)$. The
optical density $\mu(x,y)$ is then obtained from (\ref{eq:InvRadon}),
utilising the numerical scheme described in Sect. 4. The numerical
solution of (\ref{eq:RadonNum}) is obtained by using the software
Mathematica v.8. {[}12{]}

\begin{figure}
\begin{raggedright}
a) \hspace{0.45\textwidth}b)\\
\hspace{0.01\textwidth}\foreignlanguage{british}{\includegraphics[width=0.5\textwidth]{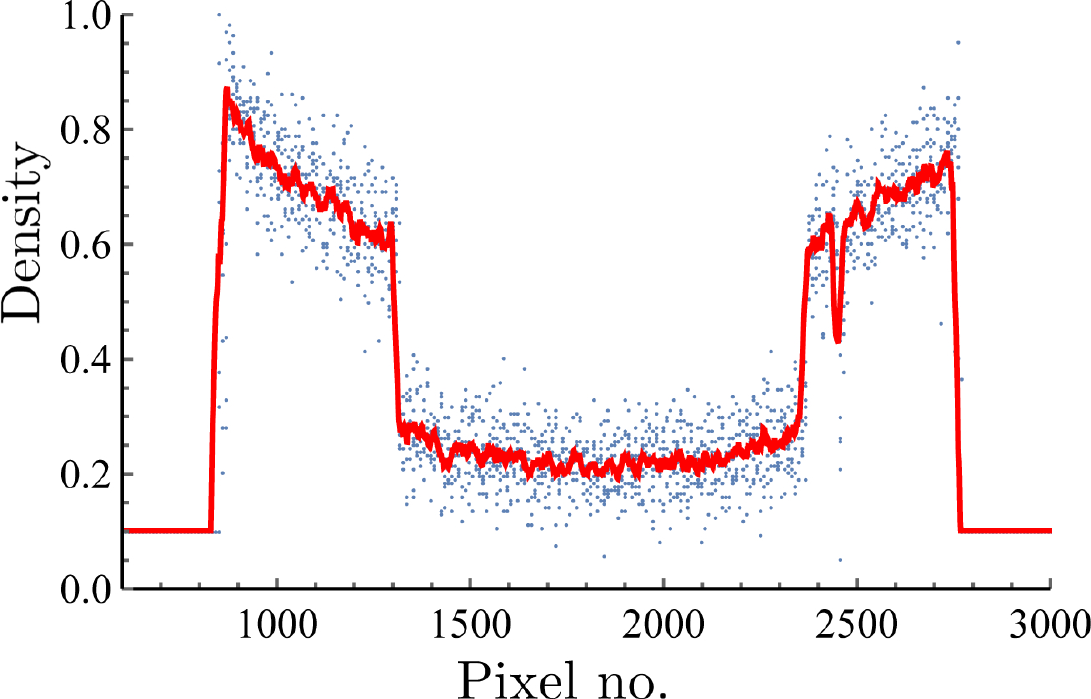}\includegraphics[width=0.5\textwidth]{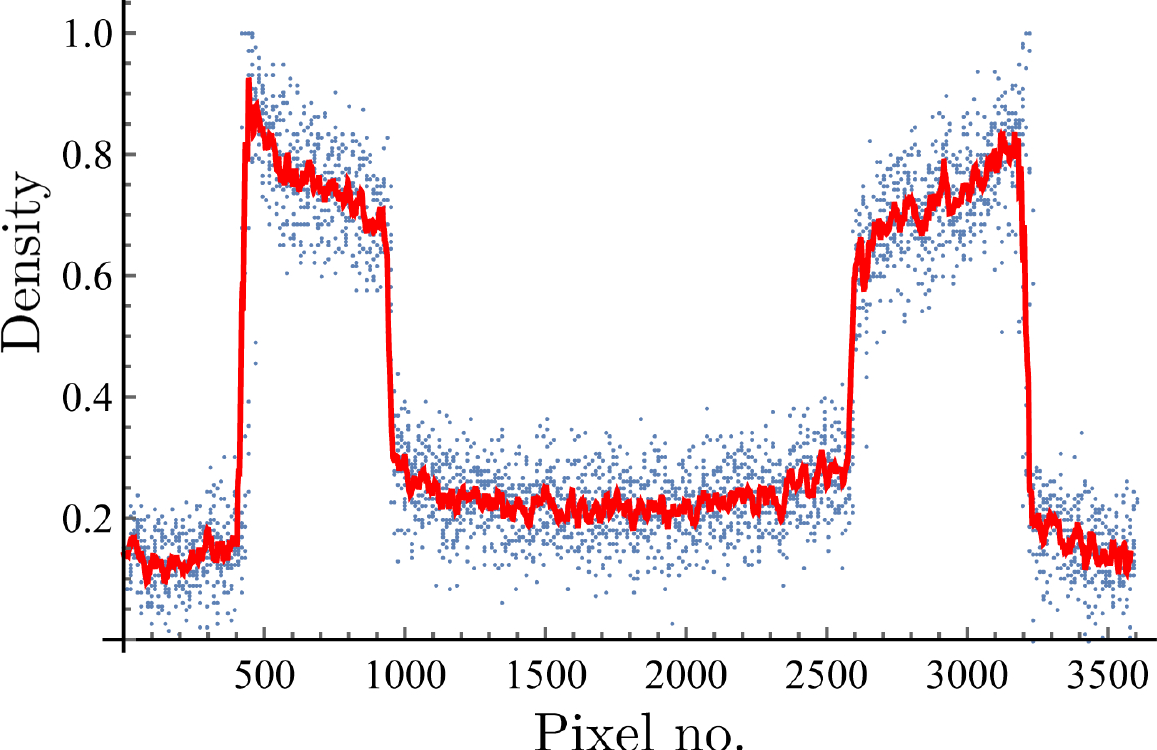}}
\par\end{raggedright}
\centering{}\label{Fig:denswscatter}\caption{Bone densities along a) the $x$-axis ($y$=0), b) the $y$-axis ($x$=0),
cf. Fig. 3. The blue dots are the actual optical densities $\mu(x,y)$
and the red curves are the filtered dittos.}
\end{figure}

Figure 4 shows the original optical density for the bone, along a)
$x=0$ and b) $y=0$, i.e. $\mu(0,y)$ and $\mu(x,0)$ respectively.
The vague swarm of blue marks is the obtained data which suffers from
a large scatter with a range of around \textpm 20\%. The data is filtered
as a moving average obtained at each current pixel as the average
of the 10 previous and 10 coming pixels. 

As observed the bone density in both directions has an expected increase
of the density towards the periosteal surface. A small but false density
is observed in the area covered by the medullary cavity that is empty.
If this is an optical effect that originates from the scanned images
or occurs as a consequence of insufficient numerical accuracy, is
at present not known.

\begin{figure}
\begin{raggedright}
a) \hspace{0.45\textwidth}b)\\
\hspace{0.01\textwidth}\foreignlanguage{british}{\includegraphics[width=0.5\textwidth]{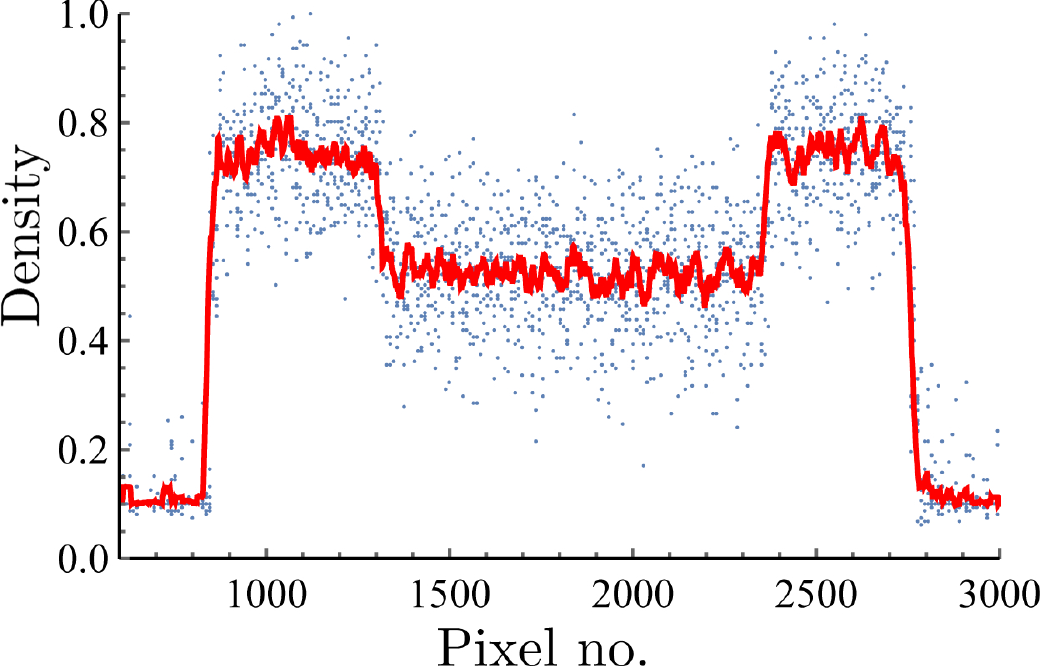}\includegraphics[width=0.5\textwidth]{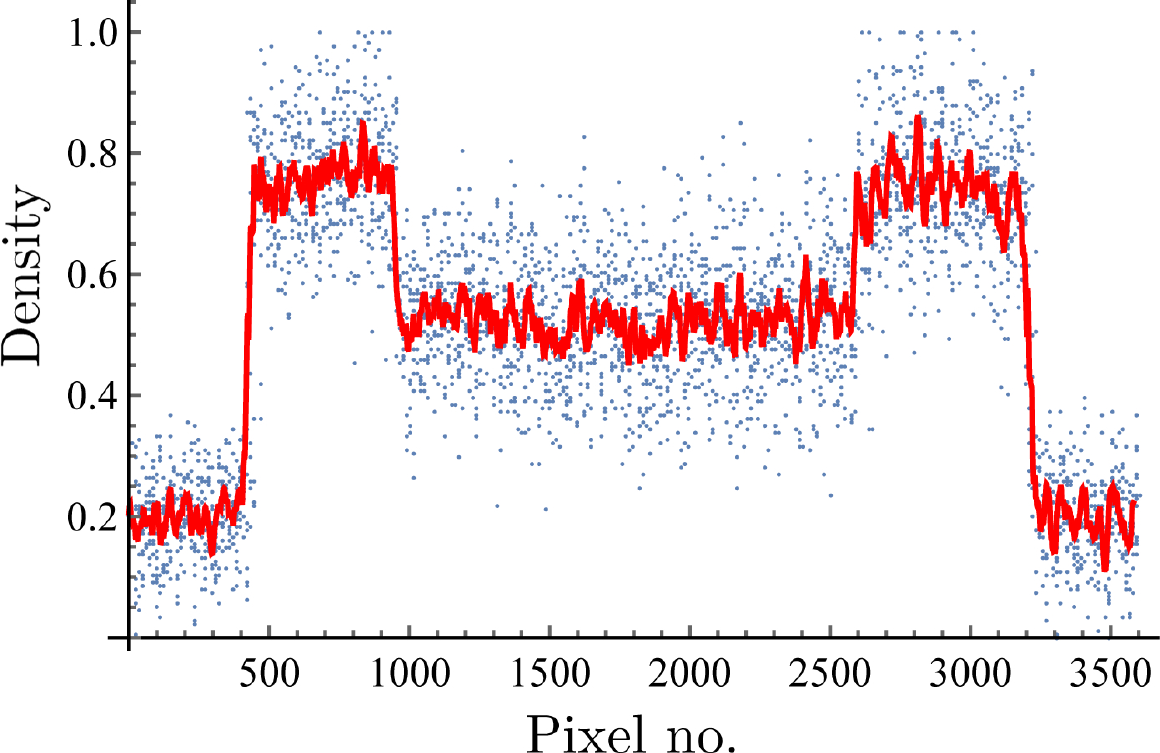}}
\par\end{raggedright}
\centering{}\label{Fig:denswscatter-iodine}\caption{Bone and iodine inserted in the medullary cavity. Densities obtained
along a) the $x$-axis, b) the $y$-axis. The pure iodine is inserted
in the medullary cavity between pixels 1000-2600 along the $x$-axis
and 1300-2800 along the $y$-axis.}
\end{figure}

In the next step iodine was inserted into the medullary cavity that
was filled only to around 60\% of it volume to avoid contamination
of the upper part of the sample. Apart from the present iodine the
procedure from the first sequence is followed. After 20 minutes a
second series of images was captured. Fig. 5 shows the results. The
striking difference is that the density inside the medullary cavity
has increased to more than 70\% of that in the bone. Also the less
dense inner parts of the bone display an increased density and reaches
that of the previously obtained density for the bone areas closest
to the periosteal surface. Visually the scatter, i.e. the blue raw
data dots, is larger and closer to around \textpm 30\%.

By taking the difference between the raw data without and with the
iodine, the distribution of the iodine is obtained. The result that
is displayed in Fig. 6 clearly shows how the iodine has penetrated
into the bone. The concentration decreases and drops to practically
zero at the periosteal surface. The distributions along $x=0$ and
$y=0$ both have the same characteristics, with the medullary cavity
having a fairly constant density and with a rapid drop at the inner
bone surface. In the bone the iodine density decreases and almost
vanishes close to the periosteal surface. This decrease may be caused
by the known increase of the bone density, leaving less space impeding
the mobility of the iodine.

\begin{figure}
\begin{raggedright}
a) \hspace{0.45\textwidth}b)\\
\hspace{0.01\textwidth}\foreignlanguage{british}{\includegraphics[scale=0.5]{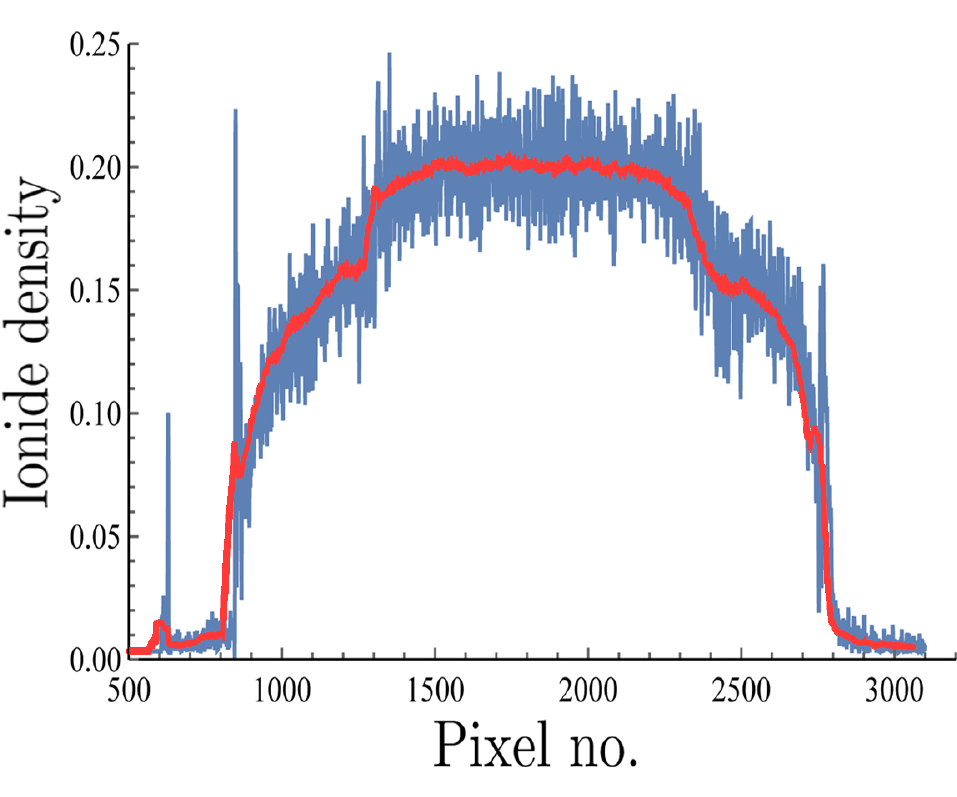}\includegraphics[scale=0.5]{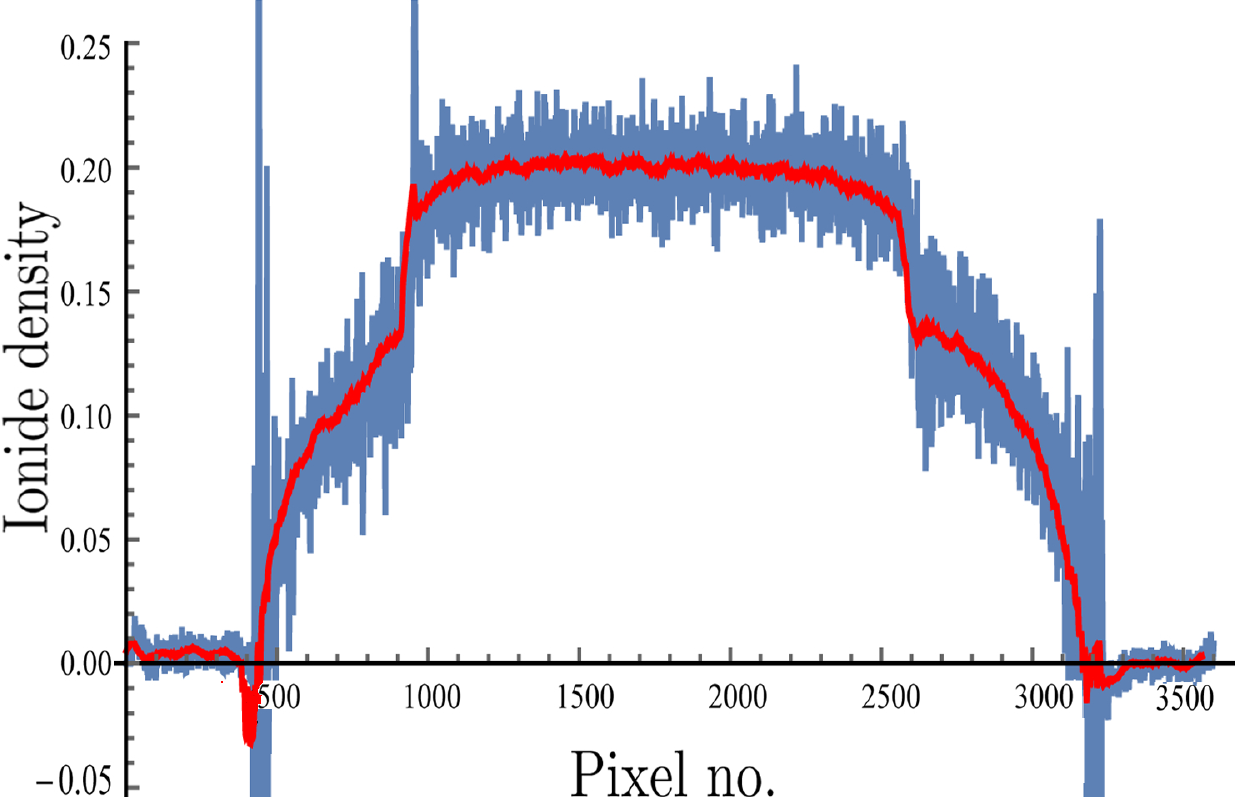}\label{Fig:comp}}
\par\end{raggedright}
\caption{The change of the optical density caused by the iodine for projections
along a) the $x$-axis, b) the $y$-axis.}
\end{figure}

Along the $x$-axis, cf. Fig. 6a), the iodine concentration decays
more rapidly in the bone section at $x<0$ where the load is compressive
and small as compared with the bone section at $x>0$, where the iodine
concentration is almost constant to around a third of the section.
Closer to the periosteal surface the concentration drops rapidly.
A preliminary explanation is that the section at $x>0$ is the large
bending stresses. These give a large tensile stress on the medullary
cavity side and compression on the periosteal side.

Along the $y$-axis, cf. Fig. 6b), the iodine concentrations in both
bone sections $y<0$ and $y>0$ are fairly similar. Further the distributions
of the iodine are close to what is observed in the bone section with
small load, i.e., at $x<0$ and $y=0$. A possible reason is that
these sections have a combination of rather thick cross sections and
are exposed to only half of the bending stresses that are present
in the section at $x>0$, $y=0$.

\section{Conclusions}

An x-ray scan of a short section of a bovine long bone was performed.
The specimen was exposed to mechanical stress via wedge forced into
a longitudinal crack. Scans were performed with and without iodine
inserted into the medullary cavity. 6284 scanned images were taken
during rotation of the specimen.

Two cross-section images before respectively after the exposure to
iodine are computed using Radon's inverse transform. The iodine penetration
into the bone was obtained as the differences between the two cross-section
images.

In conclusion the preliminary result is that the part of the cross-section
having the largest stress gradient has absorbed the highest amount
of iodine. Hence the study strengthens the hypotheses about stress-driven
diffusion in bone tissue suggested in {[}4, 5{]}. 

\section*{Acknowledgment}

The authors would like to thank The Royal Institute of Technology
for supporting G. Lindberg during a two year postdoc position. Prof.
Philip Withers, is acknowledged for allowing us to perform the experiments
at The Henry Moseley X-ray Imaging Facility, Photon Science Institute
at the University of Manchester and Prof. C. Bjerkén at Malmö University
is acknowledged for assisting at the preparation of bone samples and
during the scanning. 

\end{document}